# Geometry-guided colloidal interactions and self-tiling of elastic dipoles formed by truncated pyramid particles in liquid crystals

Bohdan Senyuk,[1] Qingkun Liu,[1] Ephraim Bililign,[1] Philip D. Nystrom,[2] and Ivan I. Smalyukh[1,2,3,4,*]

[1]*Department of Physics, University of Colorado at Boulder, Boulder, Colorado 80309, USA*
[2]*Department of Electrical, Computer, and Energy Engineering, University of Colorado at Boulder, Boulder, Colorado 80309, USA*
[3]*Liquid Crystals Materials Research Center and Materials Science and Engineering Program, University of Colorado at Boulder, Boulder, Colorado 80309, USA*
[4]*Renewable and Sustainable Energy Institute, National Renewable Energy Laboratory and University of Colorado at Boulder, Boulder, Colorado 80309, USA*

*Email: ivan.smalyukh@colorado.edu

**Abstract**

The progress of realizing colloidal structures mimicking natural forms of organization in condensed matter is inherently limited by the availability of suitable colloidal building blocks. To enable new forms of crystalline and quasicrystalline self-organization of colloids, we develop truncated pyramidal particles that form nematic elastic dipoles with long-range electrostaticlike and geometry-guided low-symmetry short-range interactions. Using a combination of nonlinear optical imaging, laser tweezers, and video microscopy, we characterize colloidal pair interactions and demonstrate unusual forms of self-tiling of these particles into crystalline, quasicrystalline, and other arrays. Our findings are explained using an electrostatics analogy along with liquid crystal elasticity and symmetry breaking considerations, potentially expanding photonic and electro-optic applications of colloids.

Comprised of building blocks like atoms, molecules, and micelles, naturally occurring and synthetic materials exist in forms of crystals, quasicrystals, and different long-range disordered but locally well-defined structural organizations [1–4]. Recent research efforts in colloidal self-assembly focus on mimicking structural motifs naturally found in molecular and atomic condensed matter [5]. Designing low-symmetry colloidal interactions to guide these different forms of organization is a challenge [5–20]. Means of diversifying colloidal self-assembly include tuning the constituent colloidal particle shape [5] or the host medium's anisotropy (e.g., by using nematic or other mesomorphic hosts) [21–25] or both [6,7,11,26–33]. The interplay of anisotropies of a liquid crystal (LC) host and colloidal building blocks, which can take forms of rods [26,32,34], polygonal platelets [6], and even knots [30], holds great potential for designing colloidal self-assembly, especially when combined with a large number of external stimuli to align, switch, and tune LCs. Recent exploration of these soft matter paradigms enabled different forms of chainlike and crystalline assembly [6,19,21,31,35] and theoretical models of LC colloidal quasicrystals [7].

In this work we extend the development of a platform for realizing crystalline, quasicrystalline, and locally ordered forms of colloidal self-organization mediated by LC elasticity that was previously exploited mainly only for high-symmetry inclusions [21,22,34–41]. By shaping particles into concave and convex polygonal truncated pyramids (PTPs) and treating them to impose perpendicular surface boundary conditions for the LC director **n** describing local average molecular alignment [1], we realize nematic colloids that interact qualitatively similar to electrostatic dipoles at large distances, but exhibit characteristic short-range interactions highly dependent on their geometric shapes. Unlike the elastic dipoles formed by spherical particles [21,22,35,41], they can be arranged into assemblies with unusual symmetries, which we demonstrate using pentagonal and rhombic PTPs. The low symmetries of the particle and the induced director field **n(r)** around it lead to long-range dipolar elastic interactions, while short-range interactions and self-tiling of the particles are guided by low-symmetry field configurations determined by their shapes. Using a combination of nonlinear optical imaging, laser tweezers, and videomicroscopy [42], we characterize colloidal pair-interaction forces and reveal unusual forms of self-tiling of interest for photonic and electro-optic applications.

Colloidal silica PTPs were fabricated using direct writing laser photolithography. A silica ($SiO_2$) layer with a thickness of 0.5–1 $\mu$m was deposited on a silicon wafer using plasma-enhanced chemical vapor deposition. Then a thin layer of photoresist AZ5214 (Clariant AG) was spin coated on the top. A lateral pattern of PTPs was defined at 405 nm using a semiconductor laser of a direct laser writing system DWL 66FS (Heidelberg Instrument), followed by inductively coupled plasma etching of PTPs into the silica layer. The isotropic chemical etching made the polygonal base of silica structure interfacing silicon slightly smaller than the one in contact with the photoresist. The photoresist was removed by acetone and the silicon substrate was etched by inductively coupled plasma, which is selective over $SiO_2$, and PTPs with rounded edges were released from the substrate. We fabricated PTPs with base planes shaped as concave or convex pentagons or rhombi, with a pyramid height of ~0.5 $\mu$m, the edge size varying within 3.5–10 $\mu$m, and the perimeter of the top and bottom pyramid bases differing by 1–2 $\mu$m. Particles were treated with an aqueous solution (~0.05 wt. %) of N,N-dimethyl-N-octadecyl-3-aminopropyl-trimethoxysilyl chloride (DMOAP) to produce normal boundary conditions for **n** and then redispersed in a nematic mixture E7 (EM Chemicals). The LC-colloidal dispersion was then filled into glass cells with a gap varied within 10–30 $\mu$m and formed by two glass substrates separated by spacers of corresponding diameter in UV-curable glue. Confining substrates were treated with DMOAP for homeotropic alignment of a far-field director **n**$_0$ or with rubbed polyimide PI2555 (HD Microsystem) to align **n**$_0$ along the rubbing direction.

   Three-photon excitation fluorescence polarizing microscopy (3PEF-PM) three-dimensional (3D) images of **n**(**r**) were obtained by multiphoton excitation of cyanobiphenyl groups of E7 molecules with a linearly polarized light at 870 nm using a tunable (680–1080 nm) Ti:sapphire oscillator (140 fs, 80 MHz, Chameleon Ultra-II, Coherent) and a setup built around an inverted microscope Olympus IX81 [42]. Liquid crystal fluorescence signals were detected within a spectral range of 387–447 nm using an optical bandpass filter and a photomultiplier tube H5784-20 (Hamamatsu). A laser scanning unit FV300 (Olympus) and a stepper motor, capable of adjusting the beam focus across the sample, were used in concert to control the 3D position of a focused excitation beam. Three-dimensional images were built from fluorescence data based on point-by-point 3D scanning using Fluoview software. Excitation light polarization was varied using a half-wave plate mounted immediately before a 100× or 60× oil immersion objective of numerical aperture ≈1.4. Videomicroscopy probed colloidal interactions [6] via recording motion of particles at rates of 15 or 60 frames per second with a CCD camera (Flea, PointGrey) and then determining their time-dependent positions from captured image sequences using ImageJ software. Optical manipulation of PTPs was realized with holographic optical tweezers operating at a wavelength of 1064 nm and integrated with the same microscope [42].

Figure 1 shows PTPs with the bases shaped as convex [Figs. 1(a) and 1(b)] and concave [Figs. 1(c) and 1(d)] pentagons in a homeotropic nematic cell. Optical [Figs. 1(a)–1(e) and 1(g)] and 3PEF-PM [Figs. 1(i)–1(l)] micrographs are consistent with homeotropic anchoring for LC molecules and $\mathbf{n}(\mathbf{r})$ at the PTP surfaces. Polygonal truncated pyramids align with their base planes normal to $\mathbf{n}_0$ and distort $\mathbf{n}(\mathbf{r})$ around their sides, with singular defect lines of strength $k = -1/2$ [1] wrapping each particle around edges of the larger-area base and closing into a loop [Figs. 1(f), 1(h), and 1(m)]. By vectorizing $\mathbf{n}(\mathbf{r})$ [43], one finds that topological hedgehog charges of the defect loop and $\mathbf{n}(\mathbf{r})$ at the PTP's surface are opposite to each other, $N = \pm 1$, consistent with topological theorems and surface Euler characteristic $\chi = 2$ determined by counting edges, faces, and vertices. Defect lines closely follow the particle sides for all shapes, including concave PTPs [Fig. 1(c)]. These defect loops localize nearby PTP edges of a base plane with the larger perimeter [Figs. 1(j) and 1(m)], at which it partially alleviates the elastic energy costs of $\mathbf{n}(\mathbf{r})$ boundary conditions at faces meeting each other at 100°–110° [1]. This defect loop localization enhances the colloidal mirror symmetry breaking with respect to particle's midplane determined by the PTP geometry, so the $\mathbf{n}(\mathbf{r})$-structure is dipolar [Fig. 1(m)]. No singular defects occur at the other edges, consistent with their rounding, <90° angles between easy axes setting boundary conditions for $\mathbf{n}$ at adjacent faces, and finite surface anchoring. By probing interactions between elastic dipoles formed by colloidal microspheres and point (small ring) defects and the ones induced by PTPs [Figs. 1(o)–1(q)], we reveal the orientation of an elastic dipole moment $\mathbf{p}$, matching what one could expect from the consideration of $\mathbf{n}(\mathbf{r})$ reconstructed from 3PEF-PM images [Figs. 1(i) and 1(j)]. In contrast to polygonal platelets with tangential anchoring [6,7,19] and odd number of sides, where an elastic dipole moment is parallel to large-area faces of platelets and orthogonal to $\mathbf{n}_0$, PTPs induce $\mathbf{p}$ normal to the base planes and along $\mathbf{n}_0$ [Fig. 1(m)]. Polygonal truncated pyramid dipoles are also structurally different from the ones formed by colloidal spheres with homeotropic anchoring accompanied by point defects [21–23,35,38,39,41] because our polygon-shaped disclination loops stretch around the bases, setting the basis for geometry-guided short-range interactions studied below. Furthermore, the point group symmetry of our pentagonal PTPs with accompanying $\mathbf{n}(\mathbf{r})$-distortions and defects is $C_{5v}$ (or, in general, $C_{mv}$ for polygonal bases with m sides, having an *m*- fold $C_m$ proper rotation axis parallel to $\mathbf{n}_0$ and $m$ mirror symmetry planes containing $C_m$), different from the $C_{\infty v}$ symmetry of elastic dipoles formed by spheres. Orientations of PTP dipoles, always along $\mathbf{n}_0$, can be mutually parallel or antiparallel (Figs. 1 and 2). Bright field micrographs of oppositely oriented dipoles [Figs. 1(a)–1(d)] are influenced by scattering from defect lines encircling the larger-perimeter base and also from refraction of light encountering asymmetric $\mathbf{n}(\mathbf{r})$-deformations. This reveals dipole orientations in videomicroscopy (Fig. 2).

Brownian motion of PTPs in a nematic host is anisotropic [6,34,44], with larger translational displacements normal to $\mathbf{n}_0$ [Fig. 1(n)]. Polygonal truncated pyramids are elastically oriented with bases orthogonal to $\mathbf{n}_0$ but are free to rotate about $\mathbf{n}_0$ (Fig. 1), as the elastic free energy of $\mathbf{n}(\mathbf{r})$ deformations around them is invariant with respect to such rotations. The ratio of self-diffusion coefficients measured for translational diffusion of convex pentagonal PTPs parallel $D_\parallel$ and perpendicular $D_\perp$ to $\mathbf{n}_0$ is $D_\parallel/D_\perp \approx 0.62$ [Fig. 1(n)], differing from $D_\parallel/D_\perp > 1$ of other known elastic dipoles [34,38]. This is caused mainly by the smaller spatial extent of aligned PTPs along $\mathbf{n}_0$ than perpendicular to $\mathbf{n}_0$, which largely determines the LC's viscous drag resistance to particle motions.

Using laser tweezers and videomicroscopy, we probe colloidal pair interactions of PTP dipoles (Figs. 2 and 3). The $\mathbf{n}(\mathbf{r})$-coronas of two elastic dipoles released from laser traps partially overlap, so the total energy depends on their mutual positions [Fig. 2(b)]. Changes of free energy arising from placing the dipoles in relative positions reducing the overall elastic distortions give rise to interaction forces. The interactions are highly anisotropic, electrostatics-like at large distances, and influenced by cell confinement and short-range forces dependent on the shapes of a particle and the accompanying $k = -1/2$ singular polygon defects. In homeotropic cells, antiparallel PTP dipoles interact attractively [Figs. 2(a)–2(f)] to self-tile into 2D side-to-side assembles in a plane perpendicular to $\mathbf{n}_0$ [Figs. 4(a)–4(g)]. Polygonal truncated pyramids are laterally separated by disclinations and regions of $\mathbf{n}(\mathbf{r})$ orthogonal to the two tilted faces of neighboring particles [Fig. 2(d)] having side surfaces of neighboring PTPs at equilibrium distances of about half of their thickness. In contrast, parallel dipoles mutually repel in the same homeotropic cells [Fig. 2(g)]. Colloidal interaction forces are balanced by the viscous drag $F_v = \sigma_f \, \partial S(t)/\partial t$ [6,22,35–38] due to low-velocity (~1 $\mu$m/s) motion of PTPs in LCs with comparably high viscosity [Fig. 2(e)] and inertia effects can be neglected, where $\sigma_f$ is an effective viscous friction coefficient. By tracking positions of individual PTPs over time with videomicroscopy, we characterize Brownian motion to find $\sigma_f$ [Fig. 1(n)] and determine the time-dependent center-to-center separation $S(t)$ of particles [Fig. 2(e)], which also yields their relative velocity $\partial S(t)/\partial t$. Polygonal truncated pyramids are kept at the same cell depth in the LC bulk due to repulsive interactions of elastic dipoles with both confining substrates, so the separation vector is always roughly orthogonal to $\mathbf{n}_0$. The force $F_{s\text{-}s}$ [inset of Fig. 2(f)] of side-to-side attractive elastic interactions, calculated using experimental $S(t)$ [Fig. 2(e)], increases with decreasing $S$, yielding the maximum of ~40 pN. The maximum force of side-to-side repulsion of parallel dipoles [Fig. 2(g)] is similar in magnitude and also drops with increasing $S$. Within a fairly wide range of separations, the observed long-range side-to-side attraction follows the power law $F_{s\text{-}s} \propto S^{-4.0\pm0.3}$ [Fig. 2(f)], which is expected for elastic dipolar interactions [21,22]. The deviation from the power law dependence of attractive forces can be explained by confinement effects [40] and strong shape anisotropy of the particles that influences short-range interactions. Measured elastic forces and the depth of corresponding potential wells $U_{s\text{-}s} - U_0 \approx 16500 k_B T$, where $k_B$ is the Boltzmann constant, $T$ is temperature, and $U_0$ is a pair potential at the beginning of measurements ($t = 0$ s), are in the range of values common for LC colloids [6,21,22,36,38]. The $C_{5v}$ symmetry of $\mathbf{n}(\mathbf{r})$ around PTPs with concave bases (Fig. 1) and the ensuing colloidal interactions between them remain qualitatively the same as for their convex counterparts, but the strength of elastic binding energy weakens due to smaller overlap of distortions near smaller side faces within their assemblies.

Elastic pair interactions of parallel PTP dipoles with the separation vector along $\mathbf{n}_0$ in planar nematic cells lead to base-to-base assemblies [Figs. 3(a)–3(d)]. Particles with bases normal to $\mathbf{n}_0$ levitate below the cell midplane [Fig. 3(c)] at a depth level determined by a balance of gravitational forces, arising from the PTP's density higher than that of LC, and elastic repulsion from both confining plates. Polygonal truncated pyramids typically have one of the base sides oriented parallel to the bottom substrate, as revealed by 3PEF-PM [Fig. 3(c)]. This lowers the original $C_{5v}$ ($C_{mv}$ in general) symmetry down to $C_s$, preserving only a mirror plane passing through the particle's center of mass orthogonally to substrates and parallel to $\mathbf{n}_0$. In addition to the PTP dipole moment along $\mathbf{n}_0$, $\mathbf{p}_\parallel$, one can define a confinement-induced dipole moment $\mathbf{p}_\perp$ orthogonal to both $\mathbf{n}_0$ and substrates (Fig. 3), but the stronger $\mathbf{p}_\parallel$ dominates interactions. Indeed, PTPs with parallel $\mathbf{p}_\parallel$ attract [Figs. 3(a)–3(d)] and those with antiparallel $\mathbf{p}_\parallel$ repel, similar to electrostatic dipoles. Confinement and gravity tend to align $\mathbf{p}_\perp$ vertically in individual colloids and their assemblies [Fig. 3(c)]. Using experimental $S(t)$ [Fig. 3(e)], one finds the maximum attractive force $F_{b\text{-}b} = 2.5$ pN and the corresponding pair potential well $U_{b\text{-}b} - U_0 \approx 5000 k_B T$, smaller than for side-to-side interactions (Fig. 2). This difference can be qualitatively understood by considering the overall reduction of $\mathbf{n}(\mathbf{r})$-deformations enabled by corresponding colloidal assemblies. Particles do not come into direct contact within the base-to-base self-assemblies but equilibrate at an intersurface distance comparable to the PTP's height, as revealed in Fig. 3(d) using 3PEF-PM. When two parallel elastic dipoles in a homeotropic cell [Fig. 3(g)] are forced to a side-to-side separation of $\sim 10\,\mu$m using tweezers, overcoming the initial repulsions, they spontaneously change vertical positions across the cell and slide over each other, switching to base-to-base attractions. Interestingly, short-range elastic interactions also cause PTP rotations, so orientations of their sides are eventually matched [Fig. 3(g)].

Polygonal truncated pyramids can form complex colloidal assemblies driven by long-range dipolar and short-range geometry-dependent interactions, which we illustrate on examples of 2D organization of convex and concave pentagonal and rhombic PTPs (Figs. 4 and 5). The geometric side-to-side self-tiling enables stable colloidal structures in forms of straight [Fig. 5(c)], bent [Figs. 4(b) and 5(d)], and branching [Fig. 4(a)] chains, closed loops [Fig. 4(c)], and vertex stars [Fig. 5(e)] of congruent particles. Releasing PTPs, one by one, from laser tweezers and the ensuing many-body interactions result in 2D lattices locally resembling crystals and quasicrystals (Fig. 4) [45]. An important property of this 2D self-tiling is that it is guided by geometry of constituent PTPs. For example, Figs. 4(f) and 4(g) show a fragment of quasicrystalline Penrose tiling [7] that symmetry permits obtaining only for pentagonal PTPs. Colloidal lattices are stable over long times and resist external perturbations because the elastic binding energy is orders of magnitude stronger than the thermal energy (Figs. 2 and 3). Once assembled, 2D lattices like the ones shown in Figs. 4 and 5 can be extended to 3D colloidal assemblies through templating that arises from base-to-base interactions [Figs. 3(a)–3(d)]. The challenge of lateral expansion of small quasicrystal-like colloidal structures [Fig. 4(f)] arises in lattice locations where attractive side-to-side interactions between opposite PTP dipoles are influenced by many-body interactions with neighboring elastic dipoles parallel to that of a PTP joining the lattice, as in the region depicted in magenta color in Fig. 4(g). One could mitigate these challenges of scaling quasicrystalline self-tiling via lowering the strength of interactions through decreasing particle size and elevating temperature close to the LC clearing point, thus reducing elastic constants and strength of interactions. The order-of-magnitude difference between side-to-side and base-to-base binding energies could allow for self-assembly of 2D crystalline and quasicrystalline templates at higher temperatures and then a template-assisted assembly of the corresponding 3D structures at low temperatures, when the base-to-base binding would strengthen enough to withstand thermal fluctuations. Such multistage assembly of 3D crystals or quasicrystals utilizing different strengths and symmetries of side-to-side and base-to-base elastic interactions may enable tunable photonic composites.

To conclude, we develop colloids formed by nematic dispersions of PTPs, which interact similarly to electrostatic dipoles at large distances. Short-range interactions between such particles depend on their geometric shape and are suitable for realizing 2D and 3D crystalline, quasicrystalline, and various locally ordered low-symmetry structures, which cannot be assembled from colloidal spheres. Since many synthetic and naturally occurring molecules have $C_{mv}$ symmetries [1,5], PTP colloids could allow for modeling their interactions and self-organization in condensed phases. Crystalline and quasicrystalline lattices of PTPs can find practical uses in photonics and electro-optics.

We acknowledge discussions with Angel Martinez and Taewoo Lee. This research was supported by the U.S. Department of Energy, Office of Basic Energy Sciences, Division of Materials Sciences and Engineering, under the Award ER46921.

**Figures**

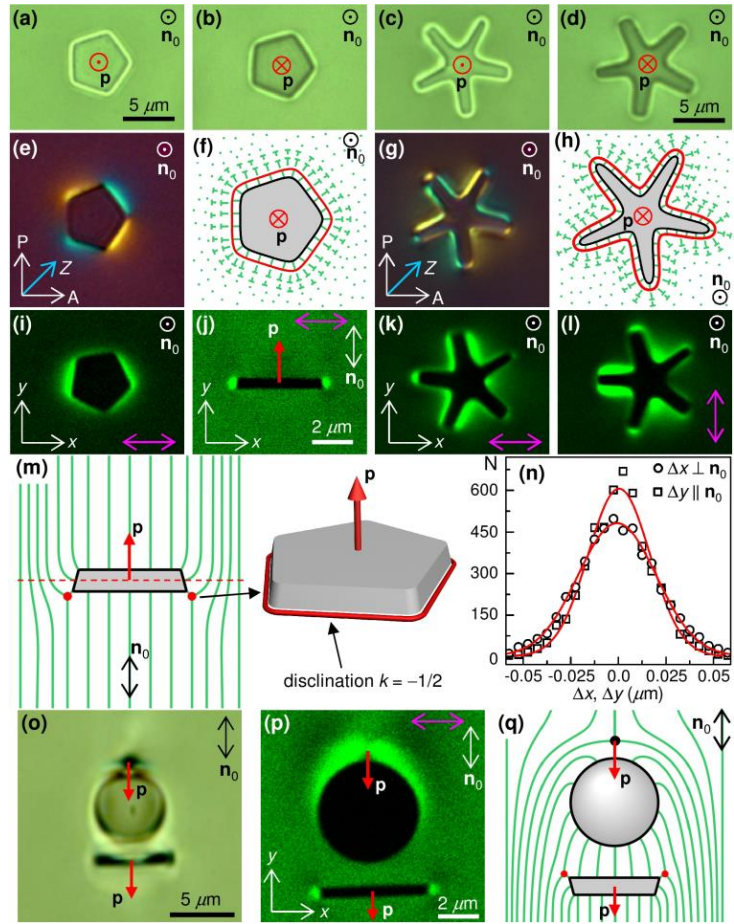

FIG. 1. Polygonal truncated pyramids in a nematic LC. (a)–(d) Bright field, (e) and (g) polarizing, and (i)–(l) 3PEF-PM images of (a) and (j) convex and (c) concave PTPs with induced **n**(**r**) distortions. Here *P*, *A*, and *Z* mark, respectively, the polarizer, analyzer and a slow axis of a 530-nm phase retardation plate. A magenta double arrow shows linear polarization of 3PEF-PM excitation light. (f), (h), and (m) Schematic of **n**(**r**) (green lines) around PTPs, which corresponds to the plane of a disclination loop (a red solid line): Nails represent molecules tilted out of the plane of an image and their length is proportional to a tilt angle; the head of the nails is below the plane of the image and their points are directed towards the observer. (n) Histograms of PTP displacements along and perpendicular to $\mathbf{n}_0$. (o) Bright field and (p) 3PEF-PM images and (q) **n**(**r**) schematic for a self-assembled dimer of dipoles formed by a microsphere and PTP.

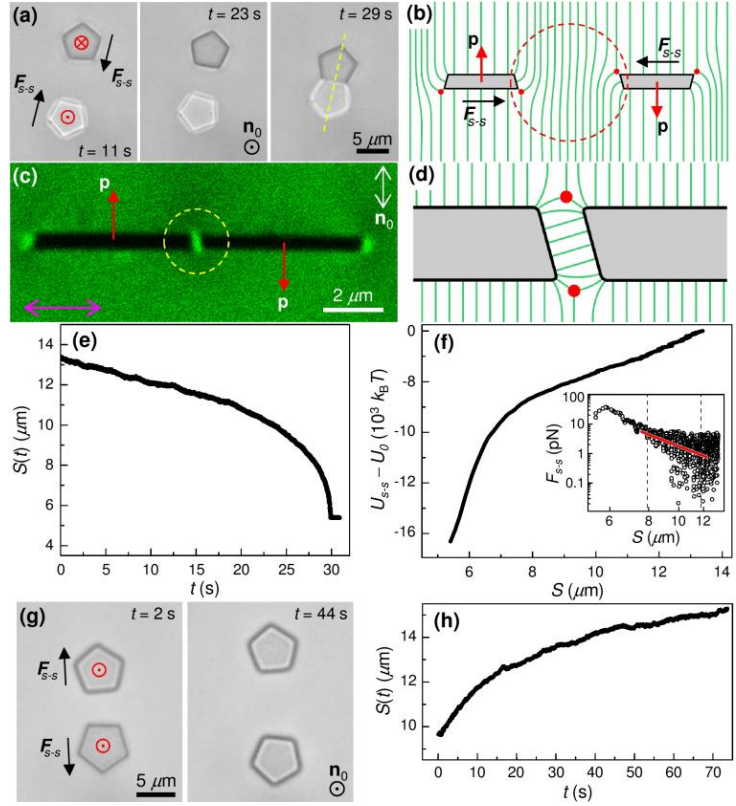

FIG. 2. Side-to-side elastic pair interactions of PTPs in a homeotropic nematic cell. (a) Sequence of optical bright field micrographs showing the attractive pair interaction between antiparallel dipoles. (b) The $\mathbf{n}(\mathbf{r})$ around interacting PTPs. (c) The 3PEF-PM cross-section obtained along a dashed yellow line shown in (a). (d) The $\mathbf{n}(\mathbf{r})$ and defects between self-assembled PTPs within the encircled area marked in (c). (e) Separation between antiparallel dipoles shown in (a) versus time. (f) The $U_{s\text{-}s} - U_0$ versus $S$; the inset shows a log-log plot of $F_{s\text{-}s}$ versus $S$ fitted (solid red line) to a power law with an exponent $-4.0 \pm 0.3$. (g) Bright field micrographs showing repulsive pair interactions between parallel dipoles. (h) The $S(t)$ for dipoles shown in (g).

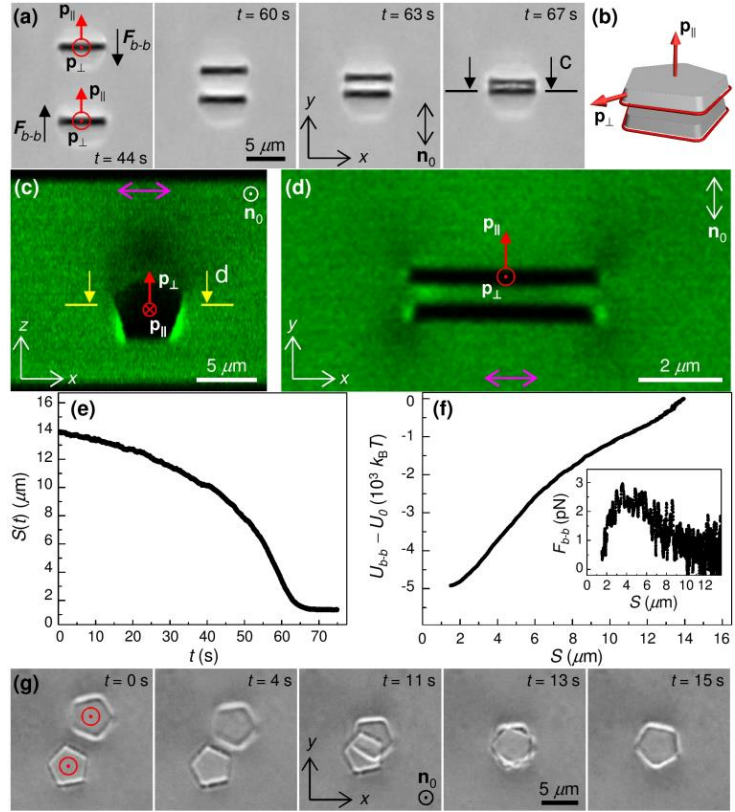

FIG. 3. (a) Bright field micrographs showing base-to-base attraction between elastic dipoles in a planar nematic cell. (b) Schematic of self-assembled PTPs and defect loops. (c) Vertical *zx* and (d) in-plane *xy* 3PEF-PM images of self-assemblies. (e) The $S(t)$ for dipoles shown in (a). (f) The $U_{b-b}$ versus $S$ for parallel elastic dipoles in (a); the inset shows the corresponding force $F_{b-b}$ versus $S$. (g) Bright field micrographs showing interactions between two elastic dipoles in a homeotropic cell when pushed towards each other; the elapsed time is marked on the images.

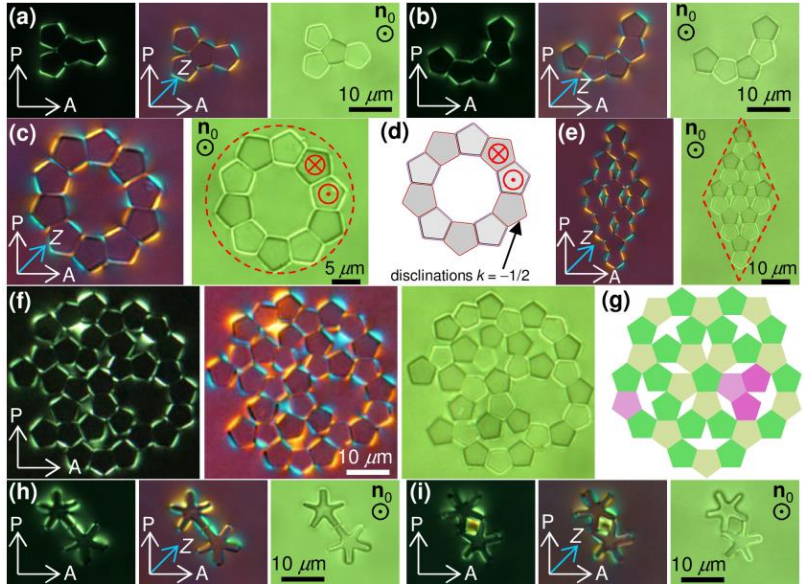

FIG. 4. Examples of assemblies of (a)–(f) convex and (h) and (i) concave pentagonal PTPs in a homeotropic nematic cell. (c) Optical images and (d) a schematic of a ringlike assembly. (e) Diamondlike assembly. (f) Micrographs of a Penrose tiling fragment and (g) the corresponding schematic; magenta pentagons represent particles missing in the experimental assembly. (h) and (i) Dimers of concave PTPs with different binding.

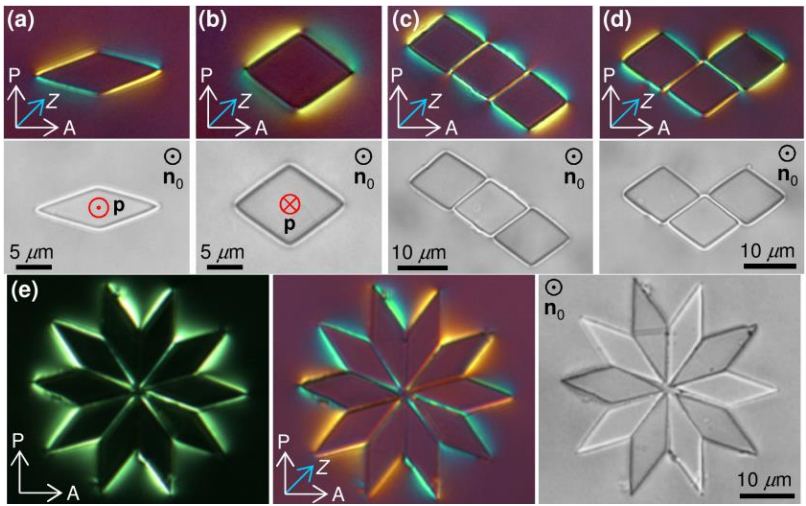

FIG. 5. Bright field and polarizing micrographs of **n**(**r**)-distortions and colloidal assemblies in a homeotropic nematic cell due to rhombic PTPs with internal angles of bases equal to (a) 36°/144° and (b) 72°/108°. (c)–(e) Examples of rhombic PTP tiling.